\pdfoutput=1

\RequirePackage{fixltx2e}
\RequirePackage{fix-cm}

\documentclass[%
floatfix,
showkeys,
nofootinbib, %
superscriptaddress, %
]{revtex4-1}

\usepackage{cmap}

\usepackage{ucs}
\usepackage[utf8x]{inputenc}
\usepackage[T1,T2A]{fontenc}
\usepackage[german,russian,english]{babel}

\usepackage[sort&compress]{natbib}
\usepackage{amsmath}
\usepackage{amssymb}

\usepackage{mathtools}
\mathtoolsset{
showonlyrefs,
mathic = true
}

\allowdisplaybreaks

\usepackage{hyperref}
\hypersetup{backref,
 colorlinks=false}
\hypersetup{pdfborder=0 0 0}

\usepackage{microtype}
\UseMicrotypeSet[protrusion]{alltext}

\usepackage{graphicx}

\usepackage[scanall]{psfrag}

\usepackage{listings}
\usepackage{listingsutf8}
\usepackage{fixltx2e}

\usepackage{nicefrac}

\makeatletter
\def\ps@pprintTitle{%
     \let\@oddhead\@empty
     \let\@evenhead\@empty
     \let\@oddfoot\@empty
     \let\@evenfoot\@oddfoot}
\makeatother

\usepackage{physics}

\newcommand{\crd}[1]{\underline{\vphantom{j}{#1}}\,}

\begin{document}

\title{The Riemannian geometry is not sufficient for the geometrization of the Maxwell's equations}

\author{T. R. Velieva}
\email{velieva_tr@rudn.university}
\affiliation{Department of Applied Probability and Informatics,\\
  Peoples' Friendship University of Russia (RUDN University),\\
  6 Miklukho-Maklaya str., Moscow, 117198, Russia}

\author{A. V. Korolkova}
\email{korolkova_av@rudn.university}
\affiliation{Department of Applied Probability and Informatics,\\
  Peoples' Friendship University of Russia (RUDN University),\\
  6 Miklukho-Maklaya str., Moscow, 117198, Russia}

\author{D. S. Kulyabov}
\email{kulyabov_ds@rudn.university}
\affiliation{Department of Applied Probability and Informatics,\\
  Peoples' Friendship University of Russia (RUDN University),\\
  6 Miklukho-Maklaya str., Moscow, 117198, Russia}
\affiliation{Laboratory of Information Technologies\\
  Joint Institute for Nuclear Research\\
  6 Joliot-Curie, Dubna, Moscow region, 141980, Russia}

\begin{abstract}
The transformation optics uses geometrized Maxwell's constitutive equations to
solve the inverse problem of optics, namely to solve the problem of
finding the parameters of the medium along the paths of
the electromagnetic field propagation.
The quadratic Riemannian geometry is usually used for the geometrization 
of Maxwell's constitutive equations, because of the usage of the general relativity
approaches.
However, the problem of the insufficiency of the
Riemannian structure for describing the constitutive tensor of the
Maxwell's equations arises.
The authors analyze the structure of the constitutive tensor and
correlate it with the structure of the metric tensor of Riemannian
geometry. It was concluded that the use of the quadratic metric for the
geometrization of Maxwell's equations is insufficient, since
the number of components of the metric tensor is less than the number
of components of the constitutive tensor.
The possible solution to this problem may be a transition to Finslerian
geometry, in particular, the use of the Berwald-Moor metric to
establish the structural correspondence between the field tensors of
the electromagnetic field.
\end{abstract}


\maketitle

\section{Introduction}
\label{sec:intro}

The problem of geometrization of Maxwell's equations arose from the
interest in Einstein's general relativity as an element of the
unified field theory.
This scientific direction was divided into two parts: the
geometrization of Maxwell's field equations and the geometrization of
Maxwell's constitutive equations~\cite{gordon:1923,tamm:1924:jrpc::en,tamm:1925:jrpc::en,tamm:1925:mathann,plebanski:1960:electromagnetic_waves,felice:1971:as_optical_medium}.
Attempts to geometrize the field equations  merged later into the gauge
field approach. The geometrization of the constitutive equations for a
long time did not find application and was almost forgotten. Interest
in geometrization was woken up by the study of metamaterials~\cite{pendry:2006:controlling-em,leonhardt:2006:mapping} 
and led to the appearance of transformation optics.

The main motivation for using the geometric approach to Maxwell's
equations is that the inverse problem of optics becomes the direct
problem of geometrized optics. The inverse problem of optics is the
problem of calculating the parameters of the medium from the known
paths of the electromagnetic waves propagation.

Transformation optics uses the geometrization of Maxwell's constitutive
equations on the basis of Riemannian geometry from the general
relativity.
However, this formalism is applicable only for a small range of
problems, when $\varepsilon_{ij} = \mu_{ij}$, that is, only
impedance matched
medium are investigated and
the parameters of the medium can be obtained only in the form of the
refractive index $n_{ij}$.

In the general case, when solving the inverse problem of
optics, we need to obtain independent permittivity and magnetic
permeability.

The structure of the paper is as follows.
In section~\ref{sec:notation} we provide the basic notation and conventions
used in the article.
In the section~\ref{sec:constitutive-tensor} the structure of
Maxwell's and Minkowski's tensors is described in the framework of the
fiber bundles. Also the structure of the constitutive tensor is
 shown. In the section~\ref{sec:tamm_geometr} the
geometrization of Maxwell's constitutive equations is described on the
basis of the Yang--Mills Lagrangian. We also did a comparison of this
geometrization with the Plebanski's geometrization. The
conclusion is drawn that the Riemannian geometry
is not sufficient for the
geometrization of Maxwell's equations.

\section{Notations and conventions}
\label{sec:notation}

  \begin{enumerate}

  \item We will use the notation of abstract
    indices~\cite{penrose-rindler:spinors::en}. In this notation tensor
    as a
    complete object is denoted merely by an index (e.g., $x^{i}$). Its
    components are
    designated by underlined indices (e.g., $x^{\crd{i}}$).
    
  \item We will adhere to the following agreements. Greek indices
    ($\alpha$, $\beta$) will refer to the four-dimensional space, in the
    component form it looks like: $\crd{\alpha} = \overline{0,3}$.  Latin
    indices from the middle of the alphabet ($i$, $j$, $k$) will refer
    to the three-dimensional space, in the component form it looks like:
    $\crd {i} = \overline{1,3}$.
    

  \item The CGS symmetrical system~\cite{sivukhin:1979:ufn::en} is used for 
  notation of the equations of electrodynamics.

  \end{enumerate}

\section{The constitutive tensor}
\label{sec:constitutive-tensor}

In what follows, we will rely on the fiber bundles theory~\cite{giachetta:2009:advanced_classical}.
The Maxwell's tensor $F_{\alpha \beta}$ is an
element of the cotangent bundle $T^{*}M$, i.e. a 2-form, and the Minkowski's tensor 
$G^{\alpha \beta}$ is an element of the tangent bundle $TM$, that is a 
bivector. We will consider the case of Riemannian geometry, so
the connection between tangents and cotangent layers is set by
effective metric $g_{\alpha \beta}$ on the basis of the bundle:
\begin{equation}
  \label{eq:diff:FGj}
  \begin{gathered}
    F = \frac{1}{2} F_{\crd{\alpha} \crd{\beta}} \dd{x^{\crd{\alpha}}}
    \wedge \dd{x^{\crd{\beta}}}, 
    \quad F \in \Lambda^{2}, \\
    G = \frac{1}{2} G^{\crd{\alpha}
      \crd{\beta}} \partial_{\crd{\alpha}}
    \wedge \partial_{\crd{\beta}}, 
    \quad
    G \in \Lambda_{2}, \\
    j = j^{\crd{\alpha}} \partial_{\crd{\alpha}},
    \quad
    j \in \Lambda_{1}.
  \end{gathered}
\end{equation}
Here $\Lambda^{2}$ is the space of 2-forms, $\Lambda_{2}$ is
the space of bivectors, $\Lambda_{1}$ is the space of vectors,
$j^{\alpha}$ is the current.

The tensors $F_{\alpha \beta}$ and $G^{\alpha \beta}$ have
the sense of curvature in the cotangent $T^{*}M$ and the tangent
$TM$ bundles. The connection between these quantities can be defined
by means of some functional $\lambda$:
\begin{equation}
\label{eq:lambda:generic}
G^{\alpha \beta} = \lambda (F_{\gamma \delta}).
\end{equation}

To clarify the relationship between the tensors $F_{\alpha \beta}$ and
$G^{\alpha \beta}$,
we will write the Maxwell equations by using the
exterior calculus formal description:
\begin{gather}
\dd{F} = 0,
\label{eq:m:form:2}
\\
\delta G = \frac{4\pi}{c} j,
\label{eq:m:form}
\end{gather}
where $c$ is the speed of light.

The Riemannian metric is usually explicitly used in the
definition of the Hodge duality operator, so we will write
down the divergence $\delta$ not through the Hodge duality operator:
\begin{equation}
  \begin{gathered}
    * : \Lambda^{k}\to \Lambda^{n-k},
    \\
    \delta = (-1)^{k} *^{-1} \dd{} *,
  \end{gathered}
\end{equation}
but through the Poincaré duality operator:
\begin{equation}
  \begin{gathered}
    \sharp : \Lambda^{k}\to \Lambda_{n-k},
    \\
    \delta = (-1)^{k} \sharp^{-1} \dd{} \sharp.
  \end{gathered}
\end{equation}

Let's write the constitutive equations as follow:
\begin{equation}
G = \lambda ( F ).
\end{equation}

Then the equation~\eqref{eq:m:form} takes the form:
\begin{equation}
  \label{eq:m:form:sharp}
  \dd{\sharp \lambda ( F )} = \frac{4\pi}{c} \sharp j.
\end{equation}

In addition, let us obtain the Hodge duality operator without an
explicit metric specification. For this we define the isomorphism:
\begin{equation}
  \label{eq:form:ast}
  \begin{gathered}
    \ast : \Lambda^{2} \to \Lambda^{2},
    \\
    \ast : F \mapsto \sharp \lambda ( F ).
  \end{gathered}
\end{equation}
Then the equation~\eqref{eq:m:form:sharp} takes the form:
\begin{equation}
  \label{eq:m:form:sharp:2}
  \dd{{}^{\ast} F} = \frac{4\pi}{c} \sharp j,
\end{equation}
and the operator~\eqref{eq:form:ast} is the Hodge duality operator,
defined not via the Riemannian metric, but through the functional $\lambda$.

By the virtue of the fact that the most practical problems consider linear media,  for simplicity we will
make the following assumptions. We will assume that the mapping
$\lambda: \Lambda^2 \to \Lambda_2$ is a linear (the connection can be
defined by means of tensors) and a local map (all tensors are considered at
the same point). Then it can be represented in the following~\cite{bpc:2::en}:
\begin{equation}
\label{eq:g_lambda_f}
G^{\alpha \beta} = 
\lambda^{\alpha \beta \gamma \delta} F_{\gamma \delta},
\end{equation}
here $\lambda^{\alpha \beta \gamma \delta}$ is constitutive tensor
that contains information about the permittivity and the 
permeability, as well as about electromagnetic constitutive
relations in Maxwell's equations~\cite{ll:8::en,tamm:1924:jrpc::en,tamm:1925:mathann}.

A nonlinear nonlocal case in the presence of translational symmetry is reduced 
to a linear local case by means of the Fourier transform. We
may write the nonlocal linear relation between $F$ and $G$ as follows:
\begin{equation}
\label{eq:lambda:linear:nonlocal}
G(x) = \int \lambda(x,s) \wedge F (s) \dd{s}, \quad
x, s \in M.
\end{equation}

Then, assuming the existence of translational invariance
$
\lambda(x,s) = \lambda(x - s),
$
we may write the connection between $F$ and $G$ in equation~\eqref{eq:lambda:linear:nonlocal}:
\begin{equation}
\label{eq:lambda:nonlinear:local:furier}
G^{\alpha \beta} (\omega, k_i) = \lambda^{\alpha \beta \gamma \delta} (\omega,k_i) F_{\gamma \delta}(\omega,k_i).
\end{equation}

From~\eqref{eq:g_lambda_f} may be seen that
$\lambda^{\alpha \beta \gamma \delta}$ has the following symmetry:
\begin{equation}
\lambda^{\alpha \beta \gamma \delta} = \lambda^{[\alpha \beta] [\gamma \delta]}.
\end{equation}

To refine the symmetry, the tensor $\lambda^{\alpha \beta \gamma
  \delta} $ can be represented in the following form~\cite{post:1972:constitutive_map,gilkey:2001:curvature_tensors,obukhov:2004:possible_skewon,hehl:2005:linear_media}:
\begin{equation}
\begin{gathered}
\lambda^{\alpha \beta \gamma \delta} =
{}^{(1)} \lambda^{\alpha \beta \gamma \delta} +
{}^{(2)} \lambda^{\alpha \beta \gamma \delta} +
{}^{(3)} \lambda^{\alpha \beta \gamma \delta},
\\
{}^{(1)} \lambda^{\alpha \beta \gamma \delta} = {}^{(1)}
\lambda^{([\alpha \beta] [\gamma \delta])}, \\
{}^{(2)} \lambda^{\alpha \beta \gamma \delta} = {}^{(2)}
\lambda^{[[\alpha \beta] [\gamma \delta]]}, \\
{}^{(3)} \lambda^{\alpha \beta \gamma \delta} = {}^{(3)}
\lambda^{[\alpha \beta \gamma \delta]}.
\end{gathered}
\end{equation}
Let's write out the number of independent components:  
\begin{itemize}
\item $\lambda^{\alpha \beta \gamma \delta}$ has 36 independent components,
\item ${}^{(1)}\lambda^{\alpha \beta \gamma \delta}$ has 20 independent components,
\item ${}^{(2)}\lambda^{\alpha \beta \gamma \delta}$ has 15 independent components,
\item ${}^{(3)}\lambda^{\alpha \beta \gamma \delta}$ has 1 independent component.
\end{itemize}

Usually only part
${}^{(1)}\lambda^{\alpha\beta\gamma\delta}$
is considered, since
${}^{(2)}\lambda^{\alpha\beta\gamma\delta}$
and
${}^{(3)}\lambda^{\alpha\beta\gamma\delta}$
make it impossible to record the electromagnetic field Lagrangian:
\begin{equation}
  \label{eq:lagrangian_em}
  L =
  - \frac{1}{16 \pi c} F_{\alpha \beta} G^{\alpha\beta} \sqrt{-g}
  - \frac{1}{c^2} A_{\alpha} j^{\alpha} \sqrt{-g}.
\end{equation}

That is, when we use parts
${}^{(2)}\lambda^{\alpha\beta\gamma\delta}$
and
${}^{(3)}\lambda^{\alpha\beta\gamma\delta}$, the tensor
$F_{\alpha \beta}$
must be self-anticommute.

\section{Geometrization of Maxwell's equations}
\label{sec:tamm_geometr}

Typically, for geometrized optics the geometrization is based on the approach proposed by
Plebanski~\cite{plebanski:1960:electromagnetic_waves,felice:1971:as_optical_medium,pendry:2006:controlling-em,leonhardt:2006:mapping}.
Briefly the program of geometrization by Plebanski can be described as
follows:
\begin{enumerate}
  \item One should write the Maxwell equations in the Minkowski space.
  \item One should write vacuum Maxwell equations in the effective Riemannian space.
  \item One should equate the corresponding members of the equations.
\end{enumerate}

As a result, we get the expression of the permittivity and 
the permeability using geometric objects, namely through the 
metric tensor of the effective Riemannian space.
This approach is somewhat similar to a mathematical trick. In addition, it still 
doesn't shed light on the actual geometrization mechanism.

The Lagrangian of the electromagnetic field~\eqref{eq:lagrangian_em} we will write in the form of a Lagrangian 
of Yang-Mills:
\begin{equation}
L = - \frac{1}{16 \pi c} g^{\alpha \gamma} g^{\beta \delta}
F_{\alpha \beta} F_{\gamma \delta} \sqrt{-g} - \frac{1}{c^2}
A_{\alpha} j^{\alpha} \sqrt{-g}.
\end{equation}
The geometrization based on Maxwell Lagrangian in the form of
Yang--Mills Lagrangian, we will call Tamm geometrization
approach~\cite{tamm:1924:jrpc::en,tamm:1925:jrpc::en,tamm:1925:mathann}.

We will construct tensor
$\lambda^{\alpha \beta \gamma \delta}$
as follows~\cite{Matagne2008}:
\begin{equation}
  \label{eq:lambda=g}
\lambda^{\alpha \beta \gamma \delta} =
2
\sqrt{-g}
g^{\alpha \beta} g^{\gamma \delta} =
\sqrt{-g}
\qty(
g^{\alpha \gamma} g^{\beta \delta} +
g^{\alpha \delta} g^{\beta \gamma}
)
+
\sqrt{-g}
\qty(
g^{\alpha \gamma} g^{\beta \delta} -
g^{\alpha \delta} g^{\beta \gamma}
).
\end{equation}

Then by taking into account the symmetry of tensors $F_{\alpha \beta}$ and $G^{\alpha \beta}$, 
equation~\eqref{eq:g_lambda_f}, with respect of~\eqref{eq:lambda=g}, will be as follows:
\begin{equation}
  G^{\alpha \beta} =
  \frac{1}{2}
  \sqrt{-g}
  \qty(
  g^{\alpha \gamma} g^{\beta \delta} -
  g^{\alpha \delta} g^{\beta \gamma}
  )
  F_{\gamma \delta}.
\end{equation}

For clarity, we will write out this equation in components:
\begin{equation}
  \label{eq:g_components}
\begin{gathered}
G^{0\crd{i}} =
\sqrt{-g}
\qty(g^{00} g^{\crd{i}\crd{j}} – g^{0\crd{i}} g^{0\crd{j}} )
F_{0\crd{j}} +
\sqrt{-g}
\qty(g^{0\crd{j}} g^{\crd{i}\crd{k}} – g^{0\crd{k}} g^{\crd{i}\crd{j}} ) F_{\crd{j}\crd{k}} ,
\\
G^{\crd{i}\crd{j}} =
\sqrt{-g}
\qty(
g^{\crd{i}0} g^{\crd{j}\crd{k}} – g^{0\crd{j}} g^{\crd{i}\crd{k}}
)
F_{0\crd{k}}
+
\sqrt{-g}
\qty(
g^{\crd{i}\crd{k}} g^{\crd{j}\crd{l}} – g^{\crd{i}\crd{l}} g^{\crd{j}\crd{k}}
)
F_{\crd{k}\crd{l}}.
\end{gathered}
\end{equation}

Let us express equations~\eqref{eq:g_components} through the field
vectors $E_{i}, B^{i}, D^{i}, H_{i}$:
\begin{gather}
  \label{eq:d^i}
  D^{\crd{i}} =
  -
  \sqrt{-g}
  \qty(g^{00} g^{\crd{i}\crd{j}} – g^{0\crd{i}} g^{0\crd{j}} )
  E_{\crd{j}} +
   \sqrt{-g}
  \varepsilon_{\crd{k} \crd{l} \crd{j}} g^{0 \crd{k}} g^{\crd{i} \crd{l}} B^{\crd{j}},
  \\
  \label{eq:b^i}
  H_{\crd{i}} =
   \sqrt{-g}
  \varepsilon_{\crd{m}\crd{n}\crd{i}} \varepsilon_{\crd{k}\crd{l}\crd{j}} g^{\crd{n}\crd{k}} g^{\crd{m}\crd{l}} B^{\crd{j}}
  +
   \sqrt{-g}
  \varepsilon^{\crd{k} \crd{l} \crd{j}} g_{0 \crd{k}} g_{\crd{i} \crd{l}} E_{\crd{j}}.
\end{gather}

From~\eqref{eq:d^i} one can formally write the expression for the
permittivity $\varepsilon^{i j}$:
\begin{equation}
  \label{eq:e_ij}
  \varepsilon^{\crd{i} \crd{j}} =
  -
   \sqrt{-g}
   \qty(g^{00} g^{\crd{i}\crd{j}} – g^{0\crd{i}} g^{0\crd{j}} )
   .
\end{equation}

  From~\eqref{eq:b^i} one can formally write the expression for the
  permeability $\mu^{i j}$:
\begin{equation}
  \label{eq:mu_ij}
  (\mu^{-1})_{\crd{i} \crd{j}} =
  \sqrt{-g}
  \varepsilon_{\crd{m}\crd{n}\crd{i}} \varepsilon_{\crd{k}\crd{l}\crd{j}} g^{\crd{n}\crd{k}} g^{\crd{m}\crd{l}}.
\end{equation}

  Thus, the geometrized constitutive equations in the components have
  the following form:
\begin{equation}
  \label{eq:geom-maxwell:tamm:decart}
  \begin{gathered}
    D^{i} = \varepsilon^{i j} E_{j} + {}^{(1)}\gamma^{i}_{j} B^{j},
    \\
    H_{i} = (\mu^{-1})_{i j} B^{j} + {}^{(2)}\gamma^{j}_{i} E_{j}, \\
    \varepsilon^{\crd{i} \crd{j}} =
    -
    \sqrt{-g}
    \qty(g^{00} g^{\crd{i}\crd{j}} – g^{0\crd{i}} g^{0\crd{j}} )
    ,
    \\
    (\mu^{-1})_{\crd{i} \crd{j}} =
    \sqrt{-g}
    \varepsilon_{\crd{m}\crd{n}\crd{i}} \varepsilon_{\crd{k}\crd{l}\crd{j}} g^{\crd{n}\crd{k}} g^{\crd{m}\crd{l}},
    \\
    {}^{(1)}\gamma^{i}_{j} = {}^{(2)}\gamma^{i}_{j}
    =
    \sqrt{-g}
    \varepsilon_{\crd{k}\crd{l}\crd{j}} g^{0\crd{k}} g^{\crd{i}\crd{l}}.
  \end{gathered}
\end{equation}

Since this geometrization and Plebanski
geometrization~\cite{plebanski:1960:electromagnetic_waves,kulyabov:2017:sfm:geometrization_maxwell}
are done on the basis of Riemannian geometry, it is possible to
demonstrate their similarity.
Indeed, it is easy to show that for Tamm's  geometrization approach the
following equation is valid
$
    \varepsilon^{\crd{i} \crd{j}} = \mu^{\crd{i} \crd{j}}
$
under the condition $g^{0 \crd{i}} = 0$.
This means that the geometrization of Maxwell's constitutive equations on
the basis of a quadratic metric imposes a restriction on the
impedance:
\begin{equation}
Z = \sqrt{\frac{\mu}{\varepsilon}} = 1.
\end{equation}

This result is a consequence of the insufficient number of components
of the Riemannian metric tensor $g_{\alpha \beta}$ (10 components),
even for tensor ${}^{(1)}\lambda^{\alpha\beta\gamma\delta}$
(20 components), not to mention the total tensor
$\lambda^{\alpha\beta\gamma\delta}$
(36 components). Even the usage of the geometrization of Riemannian geometry
with torsion and
nonmetricity~\cite{shouten:tensor_analysis::en,horsley:2011:non-riemannian}
does not change the situation.
Actually, when geometriztion is based on the Riemannian geometry, we the refractive index $n_{i j}$, but 
not the permittivity $\ varepsilon_{i j}$ and the permeability
$\mu_{i j}$.

The authors suggest that in order to solve the problem of the geometrization of the
Maxwell's equations one need to rely on Finsler geometry.
We propose to consider the equation~\eqref{eq:form:ast} as the basis for the
geometrization. As a metric, we propose to
use the Berwald-Moor
metric~\cite{rund:book:finsler::en,asanov:book:finsler::en} 
with interval as follows:
\begin{equation}
  \dd{s^4} = g_{\alpha \beta \gamma \delta}
  \dd{x^{\alpha}} \dd{x^{\beta}} \dd{x^{\gamma}} \dd{x^{\delta}}.
\end{equation}

With this choice of metric, it is possible to obtain the full number
of components for the tensor $\lambda_{\alpha \beta \gamma \delta} $.
This approach is expected to be implemented in further
research.

\section{Conclusion}
\label{sec:conclusion}

The authors demonstrated that the geometrization on
the basis of the quadratic geometry can not adequately describe the 
Maxwell's equations, since it does not allow us to investigate the
general case.
That is why, according to the authors, this direction for a long time
could not find an adequate application in practice.
In fact, it found an application only within the framework of
transformation optics for the calculation of metamaterials, since it
was required to obtain exactly the refractive index for impedance
matched materials. As an option for solving this problem, the authors
propose to use the Finsler geometry, namely the Berwald-Moor space.

\begin{acknowledgments}

The work is partially supported by Russian Foundation for Basic Research (RFBR) grants No~16-07-00556.
Also the publication was prepared with the support of the
``RUDN University Program 5-100''.

\end{acknowledgments}

 \bibliographystyle{elsarticle-num}

\bibliography{bib/finsler-maxwell/cite}

\begin{thebibliography}{10}
\expandafter\ifx\csname url\endcsname\relax
  \def\url#1{\texttt{#1}}\fi
\expandafter\ifx\csname urlprefix\endcsname\relax\def\urlprefix{URL }\fi
\expandafter\ifx\csname href\endcsname\relax
  \def\href#1#2{#2} \def\path#1{#1}\fi

\bibitem{gordon:1923}
W.~Gordon, {Zur Lichtfortpflanzung nach der Relativit{\"{a}}tstheorie}, Annalen
  der Physik 72 (1923) 421--456.
\newblock \href {http://dx.doi.org/10.1002/andp.19233772202}
  {\path{doi:10.1002/andp.19233772202}}.

\bibitem{tamm:1924:jrpc::en}
I.~E. Tamm, {Electrodynamics of an Anisotropic Medium in a Special Theory of
  Relativity}, Russian Journal of Physical and Chemical Society. Part physical
  56~(2-3) (1924) 248--262.

\bibitem{tamm:1925:jrpc::en}
I.~E. Tamm, {Crystal Optics Theory of Relativity in Connection with Geometry
  Biquadratic Forms}, Russian Journal of Physical and Chemical Society. Part
  physical 57~(3-4) (1925) 209--240.

\bibitem{tamm:1925:mathann}
I.~E. Tamm, L.~I. Mandelstam, {Elektrodynamik der anisotropen Medien in der
  speziellen Relativitatstheorie}, Mathematische Annalen 95~(1) (1925)
  154--160.

\bibitem{plebanski:1960:electromagnetic_waves}
J.~Plebanski, {Electromagnetic Waves in Gravitational Fields}, Physical Review
  118~(5) (1960) 1396--1408.
\newblock \href {http://dx.doi.org/10.1103/PhysRev.118.1396}
  {\path{doi:10.1103/PhysRev.118.1396}}.

\bibitem{felice:1971:as_optical_medium}
F.~Felice, {On the Gravitational Field Acting as an Optical Medium}, General
  Relativity and Gravitation 2~(4) (1971) 347--357.
\newblock \href {http://dx.doi.org/10.1007/BF00758153}
  {\path{doi:10.1007/BF00758153}}.

\bibitem{pendry:2006:controlling-em}
J.~B. Pendry, D.~Schurig, D.~R. Smith, {Controlling Electromagnetic Fields},
  Science 312~(5781) (2006) 1780--1782.
\newblock \href {http://dx.doi.org/10.1126/science.1125907}
  {\path{doi:10.1126/science.1125907}}.

\bibitem{leonhardt:2006:mapping}
U.~Leonhardt, {Optical Conformal Mapping}, Science 312~(June) (2006)
  1777--1780.
\newblock \href {http://arxiv.org/abs/0602092} {\path{arXiv:0602092}}, \href
  {http://dx.doi.org/10.1126/science.1218633}
  {\path{doi:10.1126/science.1218633}}.

\bibitem{penrose-rindler:spinors::en}
R.~Penrose, W.~Rindler, {Spinors and Space-Time: Volume 1, Two-Spinor Calculus
  and Relativistic Fields}, Vol.~1, Cambridge University Press, 1987.
\newblock \href {http://dx.doi.org/10.1017/CBO9780511564048}
  {\path{doi:10.1017/CBO9780511564048}}.

\bibitem{sivukhin:1979:ufn::en}
D.~V. Sivukhin, {The international system of physical units}, Soviet Physics
  Uspekhi 22~(10) (1979) 834--836.
\newblock \href {http://dx.doi.org/10.1070/PU1979v022n10ABEH005711}
  {\path{doi:10.1070/PU1979v022n10ABEH005711}}.

\bibitem{giachetta:2009:advanced_classical}
G.~Giachetta, L.~Mangiarotti, G.~A. Sardanashvily, {Advanced Classical Field
  Theory}, World Scientific Publishing Company, Singapore, 2009.
\newblock \href {http://dx.doi.org/10.1142/7189} {\path{doi:10.1142/7189}}.

\bibitem{ll:8::en}
L.~D. Landau, E.~M. Lifshitz, L.~P. Pitaevskii, {Electrodynamics of Continuous
  Media}, 2nd Edition, Course of Theoretical Physics. Vol. 8,
  Butterworth-Heinemann, 1984.

\bibitem{post:1972:constitutive_map}
E.~Post, {The constitutive map and some of its ramifications}, Annals of
  Physics 71~(2) (1972) 497--518.
\newblock \href {http://dx.doi.org/10.1016/0003-4916(72)90129-7}
  {\path{doi:10.1016/0003-4916(72)90129-7}}.

\bibitem{gilkey:2001:curvature_tensors}
P.~B. Gilkey, {Algebraic Curvature Tensors}, in: Geometric Properties of
  Natural Operators Defined by the Riemann Curvature Tensor, World Scientific
  Publishing Company, 2001, pp. 1--91.
\newblock \href {http://dx.doi.org/10.1142/9789812799692_0001}
  {\path{doi:10.1142/9789812799692_0001}}.

\bibitem{obukhov:2004:possible_skewon}
Y.~N. Obukhov, F.~W. Hehl, {Possible skewon effects on light propagation},
  Physical Review D - Particles, Fields, Gravitation and Cosmology 70~(12)
  (2004) 1--14.
\newblock \href {http://arxiv.org/abs/0409155} {\path{arXiv:0409155}}, \href
  {http://dx.doi.org/10.1103/PhysRevD.70.125015}
  {\path{doi:10.1103/PhysRevD.70.125015}}.

\bibitem{hehl:2005:linear_media}
F.~W. Hehl, Y.~N. Obukhov, {Linear media in classical electrodynamics and the
  Post constraint}, Physics Letters, Section A: General, Atomic and Solid State
  Physics 334~(4) (2005) 249--259.
\newblock \href {http://arxiv.org/abs/0411038} {\path{arXiv:0411038}}, \href
  {http://dx.doi.org/10.1016/j.physleta.2004.11.038}
  {\path{doi:10.1016/j.physleta.2004.11.038}}.

\bibitem{Matagne2008}
E.~Matagne, {Algebraic decomposition of the electromagnetic constitutive
  tensor. A step towards a pre-metric based gravitation?}, Annalen der Physik
  (Leipzig) 17~(1) (2008) 17--27.
\newblock \href {http://dx.doi.org/10.1002/andp.200710272}
  {\path{doi:10.1002/andp.200710272}}.

\bibitem{kulyabov:2017:sfm:geometrization_maxwell}
D.~S. Kulyabov, A.~V. Korolkova, L.~A. Sevastianov, M.~N. Gevorkyan, A.~V.
  Demidova, {Geometrization of Maxwell's Equations in the Construction of
  Optical Devices}, in: V.~L. Derbov, D.~E. Postnov (Eds.), Proceedings of
  SPIE. Saratov Fall Meeting 2016: Laser Physics and Photonics XVII and
  Computational Biophysics and Analysis of Biomedical Data III, Vol. 10337,
  SPIE, 2017, pp. 103370K1--7.
\newblock \href {http://dx.doi.org/10.1117/12.2267959}
  {\path{doi:10.1117/12.2267959}}.

\bibitem{shouten:tensor_analysis::en}
J.~A. Schouten, {Tensor Analysis for Physicists}, 2nd Edition, Dover Books on
  Physics, Dover Books on Physics, 2011.

\bibitem{horsley:2011:non-riemannian}
S.~A.~R. Horsley, {Transformation Optics, Isotropic Chiral Media and
  Non-Riemannian Geometry}, New Journal of Physics 13 (2011) 1--19.
\newblock \href {http://arxiv.org/abs/1101.1755} {\path{arXiv:1101.1755}},
  \href {http://dx.doi.org/10.1088/1367-2630/13/5/053053}
  {\path{doi:10.1088/1367-2630/13/5/053053}}.

\bibitem{rund:book:finsler::en}
H.~Rund, {The Differential Geometry of Finsler Spaces}, Springer Berlin
  Heidelberg, Berlin, Heidelberg, 1959.
\newblock \href {http://dx.doi.org/10.1007/978-3-642-51610-8}
  {\path{doi:10.1007/978-3-642-51610-8}}.

\bibitem{asanov:book:finsler::en}
G.~S. Asanov, {Finsler Geometry, Relativity and Gauge Theories}, 1985.
\newblock \href {http://dx.doi.org/10.1007/978-94-009-5329-1}
  {\path{doi:10.1007/978-94-009-5329-1}}.

\end{thebibliography}


\begin{thebibliography}{10}
\def\selectlanguageifdefined#1{
\expandafter\ifx\csname date#1\endcsname\relax
\else\selectlanguage{#1}\fi}
\providecommand*{\href}[2]{{\small #2}}
\providecommand*{\url}[1]{{\small #1}}
\providecommand*{\BibUrl}[1]{\url{#1}}
\providecommand{\BibAnnote}[1]{}
\providecommand*{\BibEmph}[1]{#1}
\ProvideTextCommandDefault{\cyrdash}{\iflanguage{russian}{\hbox
  to.8em{--\hss--}}{\textemdash}}
\providecommand*{\BibDash}{\ifdim\lastskip>0pt\unskip\nobreak\hskip.2em plus
  0.1em\fi
\cyrdash\hskip.2em plus 0.1em\ignorespaces}
\renewcommand{\newblock}{\ignorespaces}

\bibitem{gordon:1923}
\selectlanguageifdefined{german}
\BibEmph{Gordon~W.} {Zur Lichtfortpflanzung nach der
  Relativit{\"{a}}tstheorie}~//
  \href{http://dx.doi.org/10.1002/andp.19233772202}{\BibEmph{Annalen der
  Physik}}. \BibDash
\newblock 1923. \BibDash
\newblock {Bd.}~72. \BibDash
\newblock S.~421--456.

\bibitem{tamm:1924:jrpc::ru}
\selectlanguageifdefined{russian}
\BibEmph{Тамм~И.~Е.} {Электродинамика
  анизотропной среды в специальной теории
  относительности}~// \BibEmph{Журнал Русского
  физико-химического общества. Часть
  физическая}. \BibDash
\newblock 1924. \BibDash
\newblock Т.~56, {№} 2-3. \BibDash
\newblock {С.}~248--262.

\bibitem{tamm:1925:jrpc::ru}
\selectlanguageifdefined{russian}
\BibEmph{Тамм~И.~Е.} {Кристаллооптика теории
  относительности в связи с геометрией
  биквадратичной формы}~// \BibEmph{Журнал
  Русского физико-химического общества.
  Часть физическая}. \BibDash
\newblock 1925. \BibDash
\newblock Т.~57, {№} 3-4. \BibDash
\newblock {С.}~209--240.

\bibitem{tamm:1925:mathann}
\selectlanguageifdefined{german}
\BibEmph{Tamm~I.~E., Mandelstam~L.~I.} {Elektrodynamik der anisotropen Medien
  in der speziellen Relativitatstheorie}~// \BibEmph{Mathematische Annalen}.
  \BibDash
\newblock 1925. \BibDash
\newblock {Bd.}~95, {H.}~1. \BibDash
\newblock S.~154--160.

\bibitem{plebanski:1960:electromagnetic_waves}
\selectlanguageifdefined{english}
\BibEmph{Plebanski~J.} {Electromagnetic Waves in Gravitational Fields}~//
  \href{http://dx.doi.org/10.1103/PhysRev.118.1396}{\BibEmph{Physical Review}}.
  \BibDash
\newblock 1960. \BibDash
\newblock Vol. 118, no.~5. \BibDash
\newblock P.~1396--1408.

\bibitem{felice:1971:as_optical_medium}
\selectlanguageifdefined{english}
\BibEmph{Felice~F.} {On the Gravitational Field Acting as an Optical Medium}~//
  \href{http://dx.doi.org/10.1007/BF00758153}{\BibEmph{General Relativity and
  Gravitation}}. \BibDash
\newblock 1971. \BibDash
\newblock Vol.~2, no.~4. \BibDash
\newblock P.~347--357.

\bibitem{pendry:2006:controlling-em}
\selectlanguageifdefined{english}
\BibEmph{Pendry~J.~B., Schurig~D., Smith~D.~R.} {Controlling Electromagnetic
  Fields}~//
  \href{http://dx.doi.org/10.1126/science.1125907}{\BibEmph{Science}}. \BibDash
\newblock 2006. \BibDash
\newblock Vol. 312, no. 5781. \BibDash
\newblock P.~1780--1782.

\bibitem{leonhardt:2006:mapping}
\selectlanguageifdefined{english}
\BibEmph{Leonhardt~U.} {Optical Conformal Mapping}~//
  \href{http://dx.doi.org/10.1126/science.1218633}{\BibEmph{Science}}. \BibDash
\newblock 2006. \BibDash
\newblock Vol. 312, no. June. \BibDash
\newblock P.~1777--1780. \BibDash
\newblock 0602092.

\bibitem{penrose-rindler:spinors::ru}
\selectlanguageifdefined{russian}
\BibEmph{Пенроуз~Р., Риндлер~В.} {Спиноры и
  пространство-время. Два-спинорное
  исчисление и релятивистские поля}. \BibDash
\newblock Москва~: Мир, 1987. \BibDash
\newblock Т.~1. \BibDash
\newblock {С.}~527.

\bibitem{sivukhin:1979:ufn::ru}
\selectlanguageifdefined{russian}
\BibEmph{Сивухин~Д.~В.} {О Международной системе
  физических величин}~//
  \href{http://dx.doi.org/10.3367/UFNr.0129.197910h.0335}{\BibEmph{Успехи
  физических наук}}. \BibDash
\newblock 1979. \BibDash
\newblock Т. 129, {№}~10. \BibDash
\newblock {С.}~335--338.

\bibitem{chandrasekhar:black_hole::ru}
\selectlanguageifdefined{russian}
\BibEmph{Чандрасекар~С.} {Математическая теория
  черных дыр}. \BibDash
\newblock Москва~: Мир, 1986. \BibDash
\newblock {С.}~276+355.

\bibitem{dubrovin-novikov-fomenko}
\selectlanguageifdefined{russian}
\BibEmph{Дубровин~Б.~А., Новиков~С.~П.,
  Фоменко~А.~Т.} {Современная геометрия.
  Методы и приложения}. \BibDash
\newblock 2-е изд. {изд.} \BibDash
\newblock М.~: Наука, 1986. \BibDash
\newblock {С.}~760.

\bibitem{giachetta:2009:advanced_classical}
\selectlanguageifdefined{english}
\BibEmph{Giachetta~G., Mangiarotti~L., Sardanashvily~G.~A.}
  \href{http://dx.doi.org/10.1142/7189}{{Advanced Classical Field Theory}}.
  \BibDash
\newblock Singapore~: World Scientific Publishing Company, 2009. \BibDash
\newblock P.~382.

\bibitem{bpc:2::ru}
\selectlanguageifdefined{russian}
\BibEmph{Парселл~Э.} {Электричество и
  магнетизм}. Берклеевский курс физики. Том
  2. \BibDash
\newblock Москва~: Наука, 1971. \BibDash
\newblock {С.}~444.

\bibitem{ll:8::ru}
\selectlanguageifdefined{russian}
\BibEmph{Ландау~Л.~Д., Лифшиц~Е.~М.}
  {Электродинамика сплошных сред}.
  Теоретическая физика. Т. VIII. \BibDash
\newblock 4-е {изд.} \BibDash
\newblock Москва~: Физматлит, 2003. \BibDash
\newblock {С.}~656.

\bibitem{terletskiy-rybakov:electrodinamics}
\selectlanguageifdefined{russian}
\BibEmph{Терлецкий~Я.~П., Рыбаков~Ю.~П.}
  {Электродинамика}. \BibDash
\newblock Москва~: Высшая школа, 1990. \BibDash
\newblock {С.}~352.

\bibitem{post:1972:constitutive_map}
\selectlanguageifdefined{english}
\BibEmph{Post~E.} {The constitutive map and some of its ramifications}~//
  \href{http://dx.doi.org/10.1016/0003-4916(72)90129-7}{\BibEmph{Annals of
  Physics}}. \BibDash
\newblock 1972. \BibDash jun. \BibDash
\newblock Vol.~71, no.~2. \BibDash
\newblock P.~497--518.

\bibitem{gilkey:2001:curvature_tensors}
\selectlanguageifdefined{english}
\BibEmph{Gilkey~P.~B.}
  \href{http://dx.doi.org/10.1142/9789812799692_0001}{{Algebraic Curvature
  Tensors}}~// Geometric Properties of Natural Operators Defined by the Riemann
  Curvature Tensor. \BibDash
\newblock World Scientific Publishing Company, 2001. \BibDash nov. \BibDash
\newblock P.~1--91.

\bibitem{obukhov:2004:possible_skewon}
\selectlanguageifdefined{english}
\BibEmph{Obukhov~Y.~N., Hehl~F.~W.} {Possible skewon effects on light
  propagation}~//
  \href{http://dx.doi.org/10.1103/PhysRevD.70.125015}{\BibEmph{Physical Review
  D - Particles, Fields, Gravitation and Cosmology}}. \BibDash
\newblock 2004. \BibDash
\newblock Vol.~70, no.~12. \BibDash
\newblock P.~1--14. \BibDash
\newblock 0409155.

\bibitem{hehl:2005:linear_media}
\selectlanguageifdefined{english}
\BibEmph{Hehl~F.~W., Obukhov~Y.~N.} {Linear media in classical electrodynamics
  and the Post constraint}~//
  \href{http://dx.doi.org/10.1016/j.physleta.2004.11.038}{\BibEmph{Physics
  Letters, Section A: General, Atomic and Solid State Physics}}. \BibDash
\newblock 2005. \BibDash
\newblock Vol. 334, no.~4. \BibDash
\newblock P.~249--259. \BibDash
\newblock 0411038.

\bibitem{Matagne2008}
\selectlanguageifdefined{english}
\BibEmph{Matagne~E.} {Algebraic decomposition of the electromagnetic
  constitutive tensor. A step towards a pre-metric based gravitation?}~//
  \href{http://dx.doi.org/10.1002/andp.200710272}{\BibEmph{Annalen der Physik
  (Leipzig)}}. \BibDash
\newblock 2008. \BibDash
\newblock Vol.~17, no.~1. \BibDash
\newblock P.~17--27.

\bibitem{kulyabov:2017:sfm:geometrization_maxwell}
\selectlanguageifdefined{english}
\BibEmph{Kulyabov~D.~S., Korolkova~A.~V., Sevastianov~L.~A., Gevorkyan~M.~N.,
  Demidova~A.~V.} \href{http://dx.doi.org/10.1117/12.2267959}{{Geometrization
  of Maxwell's Equations in the Construction of Optical Devices}}~//
  Proceedings of SPIE. Saratov Fall Meeting 2016: Laser Physics and Photonics
  XVII and Computational Biophysics and Analysis of Biomedical Data III~/ Ed.\
  by\ V.~L.~Derbov, D.~E.~Postnov. \BibDash
\newblock Vol.~10337. \BibDash
\newblock SPIE, 2017. \BibDash
\newblock P.~103370K1--7.

\bibitem{shouten:tensor_analysis::ru}
\selectlanguageifdefined{russian}
\BibEmph{Схоутен~Я.~А.} {Тензорный анализ для
  физиков}. \BibDash
\newblock Москва~: Наука, Главная редакция
  физико-математической литературы, 1965.
  \BibDash
\newblock {С.}~456.

\bibitem{horsley:2011:non-riemannian}
\selectlanguageifdefined{english}
\BibEmph{Horsley~S. A.~R.} {Transformation Optics, Isotropic Chiral Media and
  Non-Riemannian Geometry}~//
  \href{http://dx.doi.org/10.1088/1367-2630/13/5/053053}{\BibEmph{New Journal
  of Physics}}. \BibDash
\newblock 2011. \BibDash
\newblock Vol.~13. \BibDash
\newblock P.~1--19. \BibDash
\newblock 1101.1755.

\bibitem{rund:book:finsler::ru}
\selectlanguageifdefined{russian}
\BibEmph{Рунд~Х.} {Дифференциальная геометрия
  финслеровых пространств}. \BibDash
\newblock Москва~: Наука, 1981. \BibDash
\newblock {С.}~504.

\bibitem{asanov:book:finsler::en}
\selectlanguageifdefined{english}
\BibEmph{Asanov~G.~S.}
  \href{http://dx.doi.org/10.1007/978-94-009-5329-1}{{Finsler Geometry,
  Relativity and Gauge Theories}}. \BibDash
\newblock 1985. \BibDash
\newblock P.~384.

\end{thebibliography}

\end{document}


\title{Недостаточность римановой геометрии при геометризации уравнений Максвелла}

\author{Т. Р. Велиева}
\email{velieva_tr@rudn.university}
\affiliation{Кафедра прикладной информатики и теории вероятностей,\\
  Российский университет дружбы народов,\\
  ул. Миклухо-Маклая, д. 6, Москва, Россия, 117198}

\author{А. В. Королькова}
\email{korolkova_av@rudn.university}
\affiliation{Кафедра прикладной информатики и теории вероятностей,\\
  Российский университет дружбы народов,\\
  ул. Миклухо-Маклая, д. 6, Москва, Россия, 117198}

\author{Д. С. Кулябов}
\email{kulyabov_ds@rudn.university}
\affiliation{Кафедра прикладной информатики и теории вероятностей,\\
  Российский университет дружбы народов,\\
  ул. Миклухо-Маклая, д. 6, Москва, Россия, 117198}
\affiliation{Лаборатория информационных технологий,\\
  Объединённый институт ядерных исследований,\\
  ул. Жолио-Кюри 6, Дубна, Московская область, Россия, 141980}

\begin{abstract}
Трансформационная оптика использует геометризованные материальные
уравнения Максвелла для решения обратной задачи оптики, а именно
для решения задачи нахождения параметров среды по траекториям распространения
электромагнитного поля.
Для геометризации материальных уравнений Максвелла обычно используют
квадратичную риманову геометрию, что
обусловлено использованием подходов общей теории относительности. 
Однако возникает вопрос о недостаточности римановой структуры для
описания тензора проницаемостей, входящего в материальные уравнения Максвелла.
Авторы анализируют структуру тензора проницаемостей и
соотносят её со структурой метрического тензора римановой
геометрии. Делается вывод о недостаточности использования квадратичной
метрики для геометризации материальных уравнений
Максвелла, поскольку количество компонент метрического тензора меньше
числа компонент тензора проницаемостей.
Возможным решением данной проблемы может являться переход к
финслеровой геометрии, в частности использование метрики 
Бервальда--Моора для установления структурного соответствия между полевыми
тензорами электромагнитного поля.
\end{abstract}

  \keywords{уравнения Максвелла, тензор проницаемостей, риманова
    геометрия, финслерова геометрия, трансформационная оптика}

\maketitle

\section{Введение}
\label{sec:intro}

Задача геометризации уравнений Максвелла возникла
на волне интереса к общей теории относительности Эйнштейна
как элемент единой теории поля.
Это научное направление 
разделилось на две части: геометризация собственно полевых уравнений
Максвелла и геометризация материальных уравнений
Максвелла~\cite{gordon:1923,tamm:1924:jrpc::ru,tamm:1925:jrpc::ru,tamm:1925:mathann,plebanski:1960:electromagnetic_waves,felice:1971:as_optical_medium}.
Попытки геометризации полевых уравнений позже влились в калибровочный
полевой подход. Геометризация же материальных уравнений долго не находила
применения и была практически забыта. Интерес к ней проснулся в
рамках исследования
метаматериалов~\cite{pendry:2006:controlling-em,leonhardt:2006:mapping},
что породило новое направление~--- трансформационную оптику.

Основным побудительным мотивом использования геометрического подхода к
уравнениям Максвелла является то, что в рамках этого подхода
обратная задача оптики становится прямой задачей геометризованной
оптики. Под обратной задачей оптики мы понимаем задачу расчёта
параметров среды по известным траекториям распространения
электромагнитного излучения.

В трансформационной оптике используется геометризация материальных
уравнений Максвелла
на основе римановой геометрии, применяемой в общей теории
относительности.
Однако этот формализм применим только для узкого круга задач, когда
$\varepsilon_{ij}=\mu_{ij}$, то есть
исследуются только среды, согласованные по импедансу и
параметры среды можно получить только в форме показателя
преломления среды $n_{ij}$.

В общем случае при решении обратной задачи оптики
нам необходимо получать независимые диэлектрическую и
магнитную проницаемости.

Структура работы следующая.
В~разделе~\ref{sec:notation} даются основные обозначения и соглашения,
применяемые в статье. В разделе~\ref{sec:constitutive-tensor}
описывается структура тензоров Максвелла и Минковского в рамках
подхода расслоенных пространств. Также показана структура тензора
проницаемостей. В разделе~\ref{sec:tamm_geometr} приводится
геометризация материальных уравнений Максвелла на основе лагранжиана
Янга--Миллса. Проводится сравнение этой геометризации с геометризацией
Плебаньского. Делается заключение о недостаточности римановой
геометрии для геометризации уравнений Максвелла.

\section{Обозначения и соглашения}
\label{sec:notation}

  \begin{enumerate}

  \item Будем использовать нотацию абстрактных
    индексов~\cite{penrose-rindler:spinors::ru}. В данной нотации тензор как
    целостный объект обозначается просто индексом (например, $x^{i}$),
    компоненты обозначаются подчёркнутым индексом (например,
    $x^{\crd{i}}$).

  \item Будем придерживаться следующих соглашений.  Греческие индексы
    ($\alpha$, $\beta$) будут относиться к четырёхмерному
    пространству и в компонентном виде будут иметь следующие значения:
    $\crd{\alpha} = \overline{0,3}$. Латинские индексы из середины
    алфавита ($i$, $j$, $k$) будут относиться к трёхмерному
    пространству и в компонентном виде будут иметь следующие значения:
    $\crd{i} = \overline{1,3}$.


  \item Для записи уравнений электродинамики в работе используется
    система СГС симметричная~\cite{sivukhin:1979:ufn::ru}.

  \end{enumerate}

\section{Тензор приницаемостей}
\label{sec:constitutive-tensor}

Рассмотрение уравнений Максвелла проводится на основе теории расслоенных
пространств~\cite{chandrasekhar:black_hole::ru,dubrovin-novikov-fomenko,giachetta:2009:advanced_classical}.
При этом тензор Максвелла $F_{\alpha \beta}$ является
элементом кокасательного расслоения $T^{*}M$, то есть 2-формой, а тензор
Минковского $G^{\alpha \beta}$
является элементом касательного расслоения $TM$, то есть
бивектором. Мы будем рассматривать случай римановой геометрии, поэтому
связь между касательным и кокасательным расслоениями задаётся
эффективной метрикой $g_{\alpha \beta}$ на базе расслоения:
\begin{equation}
  \label{eq:diff:FGj}
  \begin{gathered}
    F = \frac{1}{2} F_{\crd{\alpha} \crd{\beta}} \dd{x^{\crd{\alpha}}}
    \wedge \dd{x^{\crd{\beta}}}, 
    \quad F \in \Lambda^{2}, \\
    G = \frac{1}{2} G^{\crd{\alpha}
      \crd{\beta}} \partial_{\crd{\alpha}}
    \wedge \partial_{\crd{\beta}}, 
    \quad
    G \in \Lambda_{2}, \\
    j = j^{\crd{\alpha}} \partial_{\crd{\alpha}},
    \quad
    j \in \Lambda_{1}.
  \end{gathered}
\end{equation}
  Здесь $\Lambda^{2}$~--- пространство 2-форм, $\Lambda_{2}$~---
  пространство бивекторов,
  $\Lambda_{1}$~--- пространство векторов, $j^{\alpha}$~--- ток.

Тензоры $F_{\alpha \beta}$ и $G^{\alpha \beta}$ имеют смысл кривизны
в кокасательном $T^{*}M$ и касательном $TM$ расслоениях. Связь
между этими величинами можно задать посредством некоторого функционала $\lambda$:
\begin{equation}
\label{eq:lambda:generic}
G^{\alpha \beta} = \lambda (F_{\gamma \delta}).
\end{equation}

Чтобы прояснить связь между тензорами $F_{\alpha \beta}$ и
$G^{\alpha \beta}$, запишем уравнения Максвелла в формализме внешнего
исчисления (исчисления косых форм):
\begin{gather}
\dd{F} = 0,
\label{eq:m:form:2}
\\
\delta G = \frac{4\pi}{c} j,
\label{eq:m:form}
\end{gather}
где $c$~--- скорость света.

Поскольку при определении оператора двойственности
Ходжа обычно явно используется риманова метрика, 
будем записывать дивергенцию $\delta$ не через оператор двойственности
Ходжа,
\begin{equation}
  \begin{gathered}
    * : \Lambda^{k}\to \Lambda^{n-k},
    \\
    \delta = (-1)^{k} *^{-1} \dd{} *,
  \end{gathered}
\end{equation}
а через оператор двойственности Пуанкаре:
\begin{equation}
  \begin{gathered}
    \sharp : \Lambda^{k}\to \Lambda_{n-k},
    \\
    \delta = (-1)^{k} \sharp^{-1} \dd{} \sharp.
  \end{gathered}
\end{equation}

Запишем материальные уравнения в виде:
\begin{equation}
G = \lambda ( F ).
\end{equation}

Тогда уравнение~\eqref{eq:m:form} примет вид:
\begin{equation}
  \label{eq:m:form:sharp}
  \dd{\sharp \lambda ( F )} = \frac{4\pi}{c} \sharp j.
\end{equation}

Кроме того,
  получим оператор дуальности Ходжа без явного задания метрики.
Для этого зададим изоморфизм:
\begin{equation}
  \label{eq:form:ast}
  \begin{gathered}
    \ast : \Lambda^{2} \to \Lambda^{2},
    \\
    \ast : F \mapsto \sharp \lambda ( F ).
  \end{gathered}
\end{equation}
Тогда~\eqref{eq:m:form:sharp} примет вид:
\begin{equation}
  \label{eq:m:form:sharp:2}
  \dd{{}^{\ast} F} = \frac{4\pi}{c} \sharp j,
\end{equation}
а оператор~\eqref{eq:form:ast} будет являться оператором дуальности Ходжа,
заданным не через риманову метрику, а через функционал $\lambda$.

Поскольку большая часть практических задач рассматривает линейные
среды, то для простоты сделаем следующие допущения.
Будем считать, что отображение $\lambda : \Lambda^2 \to \Lambda_2$~---
линейное (связь можно задать посредством тензоров) и локальное (все
тензоры рассматриваются в одной и той же точке). Тогда его можно
представить в следующем
виде~\cite{bpc:2::ru}:
\begin{equation}
\label{eq:g_lambda_f}
G^{\alpha \beta} = 
\lambda^{\alpha \beta \gamma \delta} F_{\gamma \delta},
\end{equation}
здесь $\lambda^{\alpha \beta \gamma \delta}$~--- тензор
приницаемостей, содержащий информацию как об диэлектрической и магнитной
проницаемостях, так и об электромагнитной
связи в материальных уравнениях Максвелла~\cite{ll:8::ru,terletskiy-rybakov:electrodinamics,tamm:1924:jrpc::ru,tamm:1925:mathann}.

Нелинейный, нелокальный случай при наличии трансляционной
симметрии сводится к линейному локальному случаю с помощью
преобразования Фурье.
Запишем нелокальную линейную связь между $F$ и $G$ следующим образом:
\begin{equation}
\label{eq:lambda:linear:nonlocal}
G(x) = \int \lambda(x,s) \wedge F (s) \dd{s}, \quad
x, s \in M.
\end{equation}

Тогда, предполагая наличие трансляционной инвариантности
$
\lambda(x,s) = \lambda(x - s),
$
можно записать связь между $F$ и $G$ в уравнении~\eqref{eq:lambda:linear:nonlocal}:
\begin{equation}
\label{eq:lambda:nonlinear:local:furier}
G^{\alpha \beta} (\omega, k_i) = \lambda^{\alpha \beta \gamma \delta} (\omega,k_i) F_{\gamma \delta}(\omega,k_i).
\end{equation}

Из~\eqref{eq:g_lambda_f} видно, что 
$\lambda^{\alpha \beta \gamma \delta}$ имеет следующую симметрию:
\begin{equation}
\lambda^{\alpha \beta \gamma \delta} = \lambda^{[\alpha \beta] [\gamma \delta]}.
\end{equation}

Для уточнения симметрии
тензор $\lambda^{\alpha \beta \gamma \delta}$ можно представить в
следующем
виде~\cite{post:1972:constitutive_map,gilkey:2001:curvature_tensors,obukhov:2004:possible_skewon,hehl:2005:linear_media}:
\begin{equation}
\begin{gathered}
\lambda^{\alpha \beta \gamma \delta} =
{}^{(1)} \lambda^{\alpha \beta \gamma \delta} +
{}^{(2)} \lambda^{\alpha \beta \gamma \delta} +
{}^{(3)} \lambda^{\alpha \beta \gamma \delta},
\\
{}^{(1)} \lambda^{\alpha \beta \gamma \delta} = {}^{(1)}
\lambda^{([\alpha \beta] [\gamma \delta])}, \\
{}^{(2)} \lambda^{\alpha \beta \gamma \delta} = {}^{(2)}
\lambda^{[[\alpha \beta] [\gamma \delta]]}, \\
{}^{(3)} \lambda^{\alpha \beta \gamma \delta} = {}^{(3)}
\lambda^{[\alpha \beta \gamma \delta]}.
\end{gathered}
\end{equation}
Выпишем количество независимых компонент:
\begin{itemize}
\item $\lambda^{\alpha \beta \gamma \delta}$ имеет 36 независимых компонент,
\item ${}^{(1)}\lambda^{\alpha \beta \gamma \delta}$~--- 20 независимых
компонент,
\item ${}^{(2)}\lambda^{\alpha \beta \gamma \delta}$~--- 15 независимых
компонент,
\item ${}^{(3)}\lambda^{\alpha \beta \gamma \delta}$~--- 1 независимая компонента.
\end{itemize}

Обычно рассматривается только часть
${}^{(1)}\lambda^{\alpha\beta\gamma\delta}$,
поскольку  ${}^{(2)}\lambda^{\alpha\beta\gamma\delta}$ и
${}^{(3)}\lambda^{\alpha\beta\gamma\delta}$ делают невозможным запись
лагранжиана электромагнитного поля:
\begin{equation}
  \label{eq:lagrangian_em}
  L =
  - \frac{1}{16 \pi c} F_{\alpha \beta} G^{\alpha\beta} \sqrt{-g}
  - \frac{1}{c^2} A_{\alpha} j^{\alpha} \sqrt{-g}.
\end{equation}

То есть при использовании частей 
${}^{(2)}\lambda^{\alpha\beta\gamma\delta}$ и
${}^{(3)}\lambda^{\alpha\beta\gamma\delta}$ тензор $F_{\alpha \beta}$
должен антикоммутировать сам с собой.

\section{Геометризация уравнений Максвелла}
\label{sec:tamm_geometr}

Обычно для геометризованной оптики проводят геометризацию, базируясь
на подходе
Плебаньского~\cite{plebanski:1960:electromagnetic_waves,felice:1971:as_optical_medium,pendry:2006:controlling-em,leonhardt:2006:mapping}.
Кратко программу геометризации по Плебаньскому можно описать следующим
образом:
\begin{enumerate}
  \item Записать уравнения Максвелла в среде в пространстве Минковского.
  \item Записать вакуумные уравнения Максвелла в эффективном римановом
  пространстве.
  \item Приравнять соответствующие члены уравнений.
\end{enumerate}

В результате мы получим выражение диэлектрической и магнитной
проницаемостей через геометрические объекты, а именно через
метрический тензор эффективного риманового пространства.
Данный подход несколько похож на математический трюк. Кроме того, он
не даёт понимания собственно механизма геометризации.

Лагранжиан электромагнитного поля~\eqref{eq:lagrangian_em} запишем в виде лагранжиана Янга--Миллса:
\begin{equation}
L = - \frac{1}{16 \pi c} g^{\alpha \gamma} g^{\beta \delta}
F_{\alpha \beta} F_{\gamma \delta} \sqrt{-g} - \frac{1}{c^2}
A_{\alpha} j^{\alpha} \sqrt{-g}.
\end{equation}
Геометризацию, основанную на лагранжиане Максвелла в виде лагранжиана
Янга--Миллса, будем называть геометризацией
Тамма~\cite{tamm:1924:jrpc::ru,tamm:1925:jrpc::ru,tamm:1925:mathann}.

Построим тензор
$\lambda^{\alpha \beta \gamma \delta}$ следующим образом~\cite{Matagne2008}:
\begin{equation}
  \label{eq:lambda=g}
\lambda^{\alpha \beta \gamma \delta} =
2
\sqrt{-g}
g^{\alpha \beta} g^{\gamma \delta} =
\sqrt{-g}
\qty(
g^{\alpha \gamma} g^{\beta \delta} +
g^{\alpha \delta} g^{\beta \gamma}
)
+
\sqrt{-g}
\qty(
g^{\alpha \gamma} g^{\beta \delta} -
g^{\alpha \delta} g^{\beta \gamma}
).
\end{equation}

Тогда, учитывая антисимметрию тензоров $F_{\alpha \beta}$ и
$G^{\alpha \beta}$, уравнение~\eqref{eq:g_lambda_f} с учётом~\eqref{eq:lambda=g}, примет следующий вид:
\begin{equation}
  G^{\alpha \beta} =
  \frac{1}{2}
  \sqrt{-g}
  \qty(
  g^{\alpha \gamma} g^{\beta \delta} -
  g^{\alpha \delta} g^{\beta \gamma}
  )
  F_{\gamma \delta}.
\end{equation}

Для наглядности распишем по компонентам:
\begin{equation}
  \label{eq:g_components}
\begin{gathered}
G^{0\crd{i}} =
\sqrt{-g}
\qty(g^{00} g^{\crd{i}\crd{j}} – g^{0\crd{i}} g^{0\crd{j}} )
F_{0\crd{j}} +
\sqrt{-g}
\qty(g^{0\crd{j}} g^{\crd{i}\crd{k}} – g^{0\crd{k}} g^{\crd{i}\crd{j}} ) F_{\crd{j}\crd{k}} ,
\\
G^{\crd{i}\crd{j}} =
\sqrt{-g}
\qty(
g^{\crd{i}0} g^{\crd{j}\crd{k}} – g^{0\crd{j}} g^{\crd{i}\crd{k}}
)
F_{0\crd{k}}
+
\sqrt{-g}
\qty(
g^{\crd{i}\crd{k}} g^{\crd{j}\crd{l}} – g^{\crd{i}\crd{l}} g^{\crd{j}\crd{k}}
)
F_{\crd{k}\crd{l}}.
\end{gathered}
\end{equation}

Выразим уравнения~\eqref{eq:g_components} через полевые векторы
$E_{i}, B^{i}, D^{i}, H_{i}$:
\begin{gather}
  \label{eq:d^i}
  D^{\crd{i}} =
  -
  \sqrt{-g}
  \qty(g^{00} g^{\crd{i}\crd{j}} – g^{0\crd{i}} g^{0\crd{j}} )
  E_{\crd{j}} +
   \sqrt{-g}
  \varepsilon_{\crd{k} \crd{l} \crd{j}} g^{0 \crd{k}} g^{\crd{i} \crd{l}} B^{\crd{j}},
  \\
  \label{eq:b^i}
  H_{\crd{i}} =
   \sqrt{-g}
  \varepsilon_{\crd{m}\crd{n}\crd{i}} \varepsilon_{\crd{k}\crd{l}\crd{j}} g^{\crd{n}\crd{k}} g^{\crd{m}\crd{l}} B^{\crd{j}}
  +
   \sqrt{-g}
  \varepsilon^{\crd{k} \crd{l} \crd{j}} g_{0 \crd{k}} g_{\crd{i} \crd{l}} E_{\crd{j}}.
\end{gather}

Из~\eqref{eq:d^i} можно формально выписать выражение для
  диэлектрической проницаемости $\varepsilon^{i j}$:
\begin{equation}
  \label{eq:e_ij}
  \varepsilon^{\crd{i} \crd{j}} =
  -
   \sqrt{-g}
   \qty(g^{00} g^{\crd{i}\crd{j}} – g^{0\crd{i}} g^{0\crd{j}} )
   .
\end{equation}

  Из~\eqref{eq:b^i} можно формально выписать выражение для
  магнитной проницаемости $\mu^{i j}$:
\begin{equation}
  \label{eq:mu_ij}
  (\mu^{-1})_{\crd{i} \crd{j}} =
  \sqrt{-g}
  \varepsilon_{\crd{m}\crd{n}\crd{i}} \varepsilon_{\crd{k}\crd{l}\crd{j}} g^{\crd{n}\crd{k}} g^{\crd{m}\crd{l}}.
\end{equation}

  Таким образом геометризованные уравнения связи в компонентах имеют
  следующий вид:
\begin{equation}
  \label{eq:geom-maxwell:tamm:decart}
  \begin{gathered}
    D^{i} = \varepsilon^{i j} E_{j} + {}^{(1)}\gamma^{i}_{j} B^{j},
    \\
    H_{i} = (\mu^{-1})_{i j} B^{j} + {}^{(2)}\gamma^{j}_{i} E_{j}, \\
    \varepsilon^{\crd{i} \crd{j}} =
    -
    \sqrt{-g}
    \qty(g^{00} g^{\crd{i}\crd{j}} – g^{0\crd{i}} g^{0\crd{j}} )
    ,
    \\
    (\mu^{-1})_{\crd{i} \crd{j}} =
    \sqrt{-g}
    \varepsilon_{\crd{m}\crd{n}\crd{i}} \varepsilon_{\crd{k}\crd{l}\crd{j}} g^{\crd{n}\crd{k}} g^{\crd{m}\crd{l}},
    \\
    {}^{(1)}\gamma^{i}_{j} = {}^{(2)}\gamma^{i}_{j}
    =
    \sqrt{-g}
    \varepsilon_{\crd{k}\crd{l}\crd{j}} g^{0\crd{k}} g^{\crd{i}\crd{l}}.
  \end{gathered}
\end{equation}

Поскольку данная геометризация, как и геометризация
Плебаньского~\cite{plebanski:1960:electromagnetic_waves,kulyabov:2017:sfm:geometrization_maxwell},
выполнена на основе римановой геометрии, то можно продемонстрировать их совпадение.
Действительно, легко показать, что в геометризации Тамма выполняется
равенство
$
    \varepsilon^{\crd{i} \crd{j}} = \mu^{\crd{i} \crd{j}}
$
при условии $g^{0 \crd{i}} = 0$.
Это значит, что геометризация материальных уравнений Максвелла на
основе квадратичной метрики накладывает ограничение на
импеданс:
\begin{equation}
Z = \sqrt{\frac{\mu}{\varepsilon}} = 1.
\end{equation}

Такой результат
является следствием недостаточного количества компонент риманового
метрического тензора $g_{\alpha\beta}$ (10 компонент) даже для
описания тензора
${}^{(1)}\lambda^{\alpha\beta\gamma\delta}$ (20 компонент), не говоря
уже о полном тензоре $\lambda^{\alpha\beta\gamma\delta}$ (36
компонент). Даже
использование при геометризации римановой геометрии
с кручением и
неметричностью~\cite{shouten:tensor_analysis::ru,horsley:2011:non-riemannian}
не исправляет положения. Фактически, при геометризации на основе
римановой геометрии мы получает не диэлектрическую $\varepsilon_{i j}$ и
магнитную $\mu_{i j}$
проницаемости, а коэффициент преломления $n_{i j}$.

Авторы предполагают, что
для решения проблемы геометризации уравнений Максвелла
необходимо опираться на финслерову
геометрию. В качестве основы для геометризации предполагается
рассматривать соотношение~\eqref{eq:form:ast}. В качестве метрики
предлагается использовать метрику Бервальда--Моора~\cite{rund:book:finsler::ru,asanov:book:finsler::en}
с интервалом
\begin{equation}
  \dd{s^4} = g_{\alpha \beta \gamma \delta}
  \dd{x^{\alpha}} \dd{x^{\beta}} \dd{x^{\gamma}} \dd{x^{\delta}}.
\end{equation}

При таком выборе метрики возможно задать полное количество компонент
для тензора $\lambda_{\alpha \beta \gamma \delta}$.
Этот подход предполагается реализовать в дальнейших исследованиях.

\section{Заключение}
\label{sec:conclusion}

Авторами продемонстрировано, что геометризация на
основе квадратичной геометрии не может адекватно описать уравнения
Максвелла, так как не позволяет исследовать общий случай. Именно
поэтому, по мнению авторов, это направление долго не могло найти
адекватного применения на практике. Фактически оно нашло применение
только в рамках трансформационной оптики для расчёта
метаматериалов,
поскольку требовалось получить именно коэффициент преломления для
материалов, согласованных по импедансу.
В качестве варианта решения данной проблемы авторами предлагается
использование финслеровой геометрии, а именно пространства Бервальда--Моора.

\def\acknowledgmentsname{Благодарности}

\begin{acknowledgments}

Работа частично поддержана грантами РФФИ \textnumero~16-07-00556.
Также публикация подготовлена при поддержке программы РУДН <<5-100>>.

\end{acknowledgments}

  \ifthenelse{\boolean{xetex}}
  {\bibliographystyle{ugost2008l}}
  {\bibliographystyle{gost2008l}}

\bibliography{bib/finsler-maxwell/cite}